\documentclass{acm_proc_article-sp}
\usepackage{balance}
\usepackage{pgfplots}
\usepackage{tikz}
\usepackage{afterpage}
\usetikzlibrary{automata,petri,shapes,arrows,matrix}
\usepgflibrary{shapes.geometric}
\usetikzlibrary{patterns}
\usepgflibrary{patterns}
\usetikzlibrary{fit}
\usetikzlibrary{decorations.pathmorphing}
\usepackage{amssymb}
\usepackage{subfig}
\usepackage{comment}
\usepackage{float}
\usepackage{soul}
\restylefloat{figure}
\newtheorem{mydef}{Definition}

\usepackage[ruled,vlined,linesnumbered]{algorithm2e}

\definecolor{myblue5}{RGB}{198, 219, 239}
\definecolor{myblue4}{RGB}{158, 202, 225}
\definecolor{myblue3}{RGB}{107, 174, 214}
\definecolor{myblue2}{RGB}{49, 130, 189}
\definecolor{myblue1}{RGB}{8, 81, 156}
\definecolor{myblue0}{RGB}{4, 61, 126}
\definecolor{myblue}{RGB}{2, 50, 100}
\definecolor{myred1}{RGB}{239, 50, 50}
\definecolor{myred2}{RGB}{200, 20, 20}
\definecolor{myred3}{RGB}{150, 20, 20}
\definecolor{myred4}{RGB}{189, 130, 49}
\definecolor{myred5}{RGB}{156, 81, 8}
\definecolor{mygreen1}{RGB}{30, 168, 58}
\definecolor{mygreen2}{RGB}{51, 204, 102}
\definecolor{myorange}{RGB}{255, 153, 102}
\definecolor{mypurple}{RGB}{153, 153, 204}
\definecolor{mypink}{RGB}{255, 52, 102}
\definecolor{myyellow}{RGB}{255, 204, 0}

\begin{document}
\title{Cohesiveness Relationships to Empower Keyword Search on Tree Data on the Web}
\date{20 November 2011}


\numberofauthors{3}
\author
{
\alignauthor
Aggeliki Dimitriou\\
       \affaddr{School of Electrical and Computer Engineering}\\
       \affaddr{National Technical University of Athens, Greece}\\
       \email{angela@dblab.ntua.gr}
\alignauthor
Ananya Dass\\
       \affaddr{Department of Computer Science}\\
       \affaddr{New Jersey Institute of Technology, USA}\\
       \email{ad292@njit.edu}
\alignauthor
Dimitri Theodoratos\\
       \affaddr{Department of Computer Science}\\
       \affaddr{New Jersey Institute of Technology, USA}\\
       \email{dth@njit.edu}
}
\maketitle

\newlength{\hatchspread}
\newlength{\hatchthickness}
\tikzset{hatchspread/.code={\setlength{\hatchspread}{#1}},
         hatchthickness/.code={\setlength{\hatchthickness}{#1}}}
\tikzset{hatchspread=3pt,
         hatchthickness=0.4pt}
\makeatletter
\pgfdeclarepatternformonly[\tikz@pattern@color,\hatchspread,\hatchthickness]
   {custom north west lines}
   {\pgfqpoint{-2\hatchthickness}{-2\hatchthickness}}
   {\pgfqpoint{\dimexpr\hatchspread+2\hatchthickness}{\dimexpr\hatchspread+2\hatchthickness}}
   {\pgfqpoint{\hatchspread}{\hatchspread}}
   {
    \pgfsetlinewidth{\hatchthickness}
    \pgfpathmoveto{\pgfqpoint{0pt}{\hatchspread}}
    \pgfpathlineto{\pgfqpoint{\dimexpr\hatchspread+0.15pt}{-0.15pt}}
    \pgfsetstrokecolor{\tikz@pattern@color}
    \pgfusepath{stroke}
   }
\makeatother

\begin{abstract}

Keyword search has been for several years the most popular technique for retrieving information over semistructured data on the web. The reason of this unprecedented success is well known and twofold: (1) the user does not need to master a complex query language to specify her requests for data, and (2) she does not need to have any knowledge of the structure of the data sources. However, these advantages come with two drawbacks: (1) as a result of the imprecision of keyword queries, there is usually a huge number of candidate results of which only very few match the user' s intent. Unfortunately, the existing semantics are ad-hoc and they generally fail to ``guess'' the user intent. (2) As the number of keywords and the size of data grows the existing approaches do not scale satisfactorily.

In this paper, we focus on keyword search on tree data and we introduce keyword queries which can express cohesiveness relationships. Intuitively, a cohesiveness relationship on keywords indicates that the instances of these keywords in a query result should form a cohesive whole, where instances of the other keywords do not interpolate. Cohesive keyword queries allow also keyword repetition and cohesiveness relationship nesting. Most importantly, despite their increased expressiveness, they enjoy both advantages of plain keyword search. We provide formal semantics for cohesive keyword queries on tree data which ranks query results on the proximity of the keyword instances. We design a stack based algorithm which builds a lattice of keyword partitions to efficiently compute keyword queries and further leverages cohesiveness relationships to significantly reduce the dimensionality of the lattice. We implemented our approach and ran extensive experiments to measure the effectiveness of keyword queries and the efficiency and scalability of our algorithm. Our results demonstrate that our approach outperforms previous filtering semantics and our algorithm scales smoothly achieving interactive response times on queries of 20 frequent keywords on large datasets.

\end{abstract}

\section{Introduction}
Currently, huge amount of data are exported and exchanged in tree-structure form \cite{DBLP:journals/tods/MozafariZDZ13,DBLP:journals/pvldb/OgdenTP13}. Tree structures can conveniently represent data that are semistructured, incomplete and irregular as is usually the case with data on the web. Keyword search has been by far the most popular technique for retrieving data on the web. There are two main reasons for this popularity: (a) The user does not need to master a complex query language (e.g., XQuery), and (b) the user does not need to have  knowledge of the schema of the data sources over which the query is evaluated. In fact, in most cases the user does not even know which are these data sources. In addition, the same keyword query can be used to extract data from multiple sources with different structure and schema which relate the keywords in different ways.

There is a price to pay for the simplicity and convenience of keyword search. Keyword queries are imprecise. As a consequence, there is usually a large number of results from which very few are relevant to the user intent. This leads to at least two drawbacks: (a) because of the large number of candidate results, previous algorithms for keyword search are of high complexity and they cannot scale satisfactorily when the number of keywords and the size of the input dataset increase, and (b) correctly identifying the relevant results among the plethora of candidate results, is a very difficult task. Indeed, it is practically impossible for a search system to ``guess'' the user intent from a keyword query and the structure of the data source. Multiple approaches assign semantics to keyword queries by exploiting structural and semantic features of the data in order to automatically filter out irrelevant results. Examples include ELCA (Exclusive LCA) \cite{DBLP:conf/sigmod/GuoSBS03,DBLP:conf/edbt/XuP08,DBLP:conf/edbt/ZhouLL10}, VLCA (Valuable LCA) \cite{DBLP:conf/vldb/CohenMKS03,DBLP:conf/cikm/LiFWZ07}, CVLCA (Compact Valuable LCA) \cite{DBLP:conf/cikm/LiFWZ07}, SLCA (Smallest LCA) \cite{DBLP:conf/icde/HristidisPB03,DBLP:conf/sigmod/XuP05,DBLP:conf/www/SunCG07,DBLP:conf/icde/ChenP10}, MaxMatch \cite{DBLP:journals/pvldb/LiuC08}, MLCA (Meaningful LCA) \cite{DBLP:conf/vldb/LiYJ04,DBLP:journals/tods/TermehchyW11}, and Schema level SLCA \cite{DBLP:conf/vldb/GoldmanW97}. A survey of some of these approaches can be found in \cite{DBLP:journals/www/LiuC11}. Although filtering approaches are intuitively reasonable, they are sufficiently ad-hoc and they are frequently violated in practice resulting in low precision and/or recall. A better technique is adopted by other approaches which rank the candidate results in descending order of their estimated relevance. Given that users are typically interested in a small number of query results, some of these approaches combine ranking with top-k algorithms for keyword search. The ranking is performed using a scoring function and is usually based on IR-style metrics for flat documents (e.g., tf*idf or PageRank) adapted to the tree-structure form of the data \cite{DBLP:conf/sigmod/GuoSBS03,DBLP:conf/vldb/CohenMKS03,DBLP:journals/sigmod/Amer-YahiaL06,DBLP:conf/icde/BaoLCL09,DBLP:journals/isci/LiLFZ09,DBLP:conf/icde/ChenP10,DBLP:journals/tods/TermehchyW11,DBLP:conf/icde/LiLZW11,DBLP:journals/www/LiuC11,DBLP:journals/www/NguyenC12}. Nevertheless, scoring functions, keyword occurrence statistics and probabilistic models alone cannot capture effectively the intent of the user. As a consequence the produced rankings are, in general, of low quality \cite{DBLP:journals/tods/TermehchyW11}.

\vspace*{.5ex}\noindent{\bf Our approach.} In this paper, we introduce a novel approach which allows the user to specify cohesiveness relationships among keywords in a query, an option not offered by the current keyword query languages. Cohesiveness relationships group together keywords in a query to define indivisible (cohesive) collections of keywords. They partially relieve the system from guessing without affecting the user who is naturally tempted to specify such relationships. 





For instance, consider the keyword query {\tt \{XML John Smith George Brown\}} to be issued against a large bibliographic database. The user is looking for publications on \texttt{XML} related to the authors \texttt{John Smith} and \texttt{George Brown}. If the user can express the fact that the matches of {\tt John} and {\tt Smith} should form a unit where the matches of the other keywords of the query \texttt{George}, \texttt{Brown} and \texttt{XML} cannot slip into (that is, the matches of \texttt{John} and \texttt{Smith} form a {\it cohesive whole}), the system would be able to return more accurate results. For example, it will be able to filter out publications on \texttt{XML} by \texttt{John Brown} and \texttt{George Smith}. It will also filter out a publication which cites a paper authored by \texttt{John} Davis, a report authored by \texttt{George Brown} and a book on \texttt{XML} authored by Tom \texttt{Smith}. This information is irrelevant and no one of the previous filtering approaches (e.g., ELCA, VLCA, CVLCA, SLCA, MLCA, MaxMach etc.) is able to automatically exclude it from the answer of the query. Furthermore, specifying cohesiveness relationships prevents wasting precious time searching for these irrelevant results. We specify cohesiveness relationships among keywords by enclosing them between parentheses. For example, the previous keyword query would be expressed as {\tt (XML (John\,Smith) (George\,Brown))}.



Note that specifying a cohesiveness relationship on a set of keywords is different than phrase matching over flat text documents in IR. For instance, Google allows a user to specify between quotes a phrase which has to be matched intact against the text document. In contrast, specifying a cohesiveness relationship on a number of keywords to be evaluated over a data tree does not impose any syntactic restriction (e.g., distance restriction) on the matches of these keywords on the data tree. It only requires that the matches of these keywords form a cohesive unit. We provide formal semantics for queries with cohesiveness relationships on tree data in Section 2.


\begin{figure*}[t]
\centering
\includegraphics[width=1.0\textwidth, keepaspectratio=true]{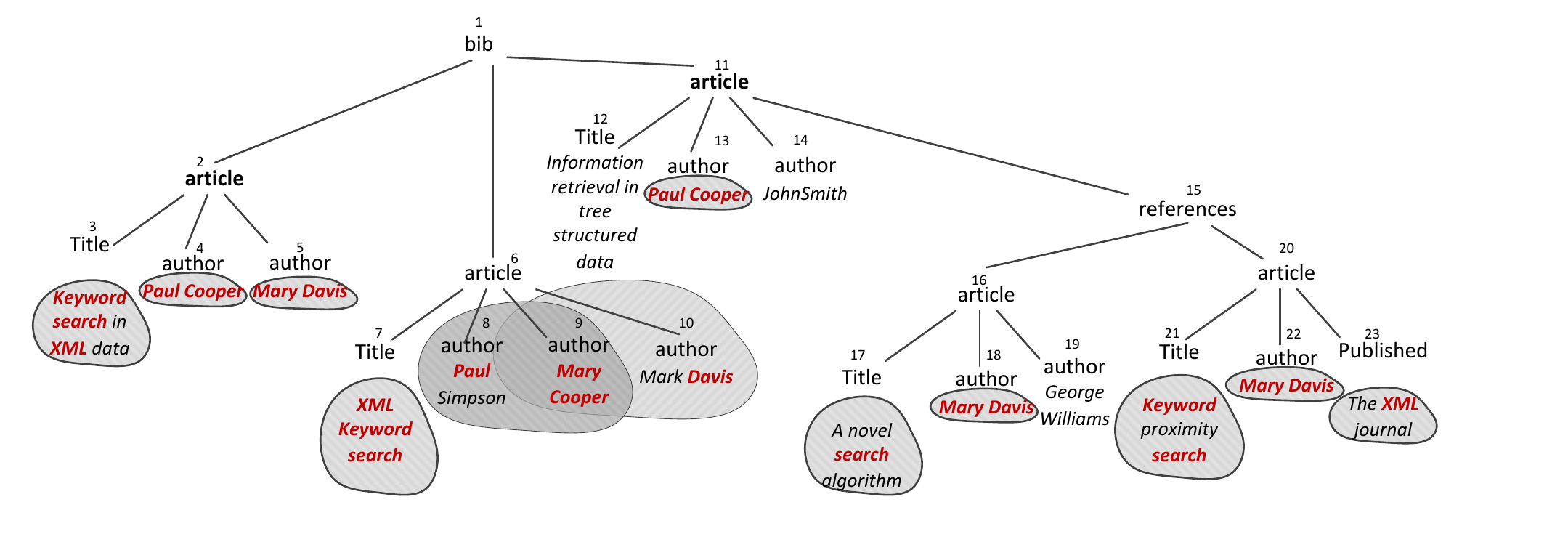}
\caption{Example data tree $D_1$}
\label{fig:exampletree2}
\end{figure*}

Cohesiveness relationships can be nested. For instance the query \texttt{(XML (John\,Smith) (citation (George\,Brown)))} looks for a paper on \texttt{XML} by \texttt{John Smith} which cites a paper by \texttt{George Brown}. The cohesive keyword query language conveniently allows also for keyword repetition. For instance, the query \texttt{(XML (John\,Smith) (citation (John\,Brown)))} looks for a paper on \texttt{XML} by \texttt{John Smith} which cites a paper by \texttt{John Brown}.

Most importantly, despite its increased expressive power, the cohesive keyword query language enjoys of both advantages of traditional keyword search. Keyword queries with cohesiveness relationships do not require any previous knowledge of a language or of the schema of the data sources and they do not require any effort by the user. The users are naturally tempted to express cohesiveness relationships when writing a keyword query and they can do it effortlessly with a cohesive query language. The benefits though in query answer quality compared to other flat approaches and performance are impressive.

\vspace*{1ex}\noindent{\bf Contribution.} The main contributions of our paper are as follows:
\begin{list}{}{\setlength{\leftmargin}{.4cm}\setlength{\parsep}{0cm}
\setlength{\partopsep}{0cm}\setlength{\itemsep}{0.1cm}\setlength{\parskip}{0cm}
\setlength{\labelwidth}{.4cm}
\setlength{\topsep}{0cm}}
\item[$\bullet$] We formally introduce a novel keyword query language which allows for cohesiveness relationships, cohesiveness relationship nesting and keyword repetition. 
\item[$\bullet$] We provide ranking semantics for the cohesive keyword queries on tree data based on the concept of LCA size \cite{DBLP:journals/is/DimitriouTS15,DBLP:conf/sigmod/DimitriouT12}. The LCA size reflects the proximity of keywords in the data tree and, similarly to keyword proximity in IR, it is used to determine the relevance of the query results.  
\item[$\bullet$] Our semantics interpret the subtrees rooted at the LCA of the instances of cohesively related keywords in the data tree as black boxes where the instances of the other keywords cannot interpolate.
\item[$\bullet$] We design an efficient multi-stack based algorithm which exploits a lattice of stacks---each stack corresponding to a different partition of the query keywords. Our algorithm does not rely on auxiliary index structures and, therefore, can be exploited on datasets which have not been preprocessed. 
\item[$\bullet$] We show how cohesive relationships can be leveraged to lower the dimensionality of the lattice and dramatically reduce its size and improve the performance of the algorithm. 
\item[$\bullet$] We analyze our algorithm and show that for a constant number of keywords it is linear on the size of the input keywords' inverted lists, i.e., to the dataset size.
Our analysis further shows that the performance of our algorithm essentially depends on the maximum cardinality of the largest cohesive term in the keyword query. 
\item{$\bullet$} We run extensive experiments on different real and benchmark datasets to assess the effectiveness of our approach and the efficiency and scalability of our algorithm. Our results show that our approach largely outperforms previous filtering approaches achieving in most cases perfect precision and recall. They also show that our algorithm scales smoothly when the number of keywords and the size of the dataset increase achieving interactive response times even with queries of 20 keywords having in total several thousands of instances on large datasets.

\end{list}

\section{Data and query model}

We consider data modeled as an ordered labeled tree. Tree nodes can represent XML elements or attributes. Every node has an id, a label (corresponding to an element tag or attribute name) and possibly a value (corresponding to the text content of an element or to an attribute's value). For identifying tree nodes we adopt the Dewey encoding scheme \cite{dewey}, which encodes tree nodes according to a preorder traversal of the data tree. The Dewey encoding scheme naturally expresses ancestor-descendant and parent-child relationships among tree nodes and conveniently supports the processing of nodes in stacks~\cite{DBLP:conf/sigmod/GuoSBS03}.

A keyword $k$ may appear in the label or in the value of a node $n$ in the data tree one or multiple times, in which case we say that node $n$ constitutes an \emph{instance} of $k$. A node may contain multiple distinct keywords in its value and label, in which case it is an instance of multiple keywords. 

A {\em cohesive keyword query} is a keyword query, which besides keywords may also contain groups of keywords called terms. Intuitively, a term expresses a cohesiveness relationship on keywords and/or terms. More formally, a keyword query is recursively defined as follows:

\begin{mydef}[Cohesive keyword query]
A \emph{term} is a multiset of at least two keywords and/or terms. A \emph{cohesive keyword query} is: (a) a set of a single keyword, or (b)~a~term. Sets and multisets are delimited within a query using parentheses.
\label{querydef}
\end{mydef}

For instance, the expression \texttt{((title XML)\,((John Smith) author))} is a keyword query. Some of its terms are \texttt{$T_1$ = (title XML)}, $T_2$ = \texttt{((John Smith) author)}, $T_3$ = \texttt{(John Smith)}, and $T_3$ is \emph{nested} into $T_2$. 

A keyword may occur multiple times in a query. For instance, in the keyword query \texttt{((journal (Information Systems) ((Information Retrieval) Smith))} the keyword \texttt{Information} occurs twice, once in the term \texttt{(Information Systems)} and once in the term \texttt{(Information Retrieval)}.

In the rest of the paper, we may refer to a cohesive keyword query simply as query. The syntax of a query $Q$ is defined by the following grammar where the non-terminal symbol $T$ denotes a term and the terminal symbol $k$ denotes a keyword:

\begin{algorithm}[!h]
\SetKwBlock{simpleblock}{}{}
\SetKwComment{mycomment}{/* }{ */}
\SetAlFnt{\tiny\sf}
\SetNoFillComment
\scriptsize
\small
\textbf{CohesiveLCA}($Q$: cohesive keyword query, $invL$: inverted lists)
\simpleblock{
    buildLattice()\\
    \While{currentNode $\gets$ getNextNodeFromInvertedLists()}
    {
		currentPLCA $\gets$ createPartialLCA(currentNode, 0, null)\\	
		push(initStack, currentPLCA)\\
        \For{every coarsenessLevel}
        {
            \While{partialLCA $\gets$ next partial LCA of this coarsenessLevel}
            {
                \For{every stack of coarsenessLevel containing termOf(partialLCA)}
                {push(stack, partialLCA)}
            }
       }
    }
    emptyStacks()\mycomment{pop all entries from stacks sequentially from each coarseness level, letting subsequent levels process newly constructed partial LCAs}
}

\textbf{push}(stack, partialLCA)
\simpleblock{
    \While{stack.dewey not ancestor of pqartialLCA.node}
    {
        pop(stack)
    }
    \While{stack.dewey $\neq$ pqartialLCA.node}
    {
        addEmptyRow(stack) \mycomment{updating stack.dewey until it is equal to pqartialLCA.node}
    }
    replaceIfSmallerWith(stack.topRow, partialLCA.term, size)
}
\textbf{pop}(stack)
\simpleblock{
   poppedEntry $\gets$ stack.pop()\\
    \If{stack.columns = 1}
    { 
        addResult(stack.dewey, popped[0].size)\\
    }
    \mycomment{Produce new LCAs from two partial LCAs}
    \If{stack.columns $>$ 1}
    {
       \For{i$\gets$0 to stack.columns}
       {
            \For{j$\gets$i to stack.columns}
            {
                \If{popped[i] and popped[j] contain sizes \textbf{and} popped[i].provenance $\cap$ popped[j].provenance~=~$\emptyset$ }
                {
                    newTerm $\gets$ findTerm(popped[i].term, popped[j]term)\\
                    newSize $\gets$ popped[i].size+popped[j].size\\
                    newProvenance $\gets$ popped[i].provenance $\cup$ popped[j].provenance \\
                    pLCA $\gets$ newPartialLCA(stack.dewey, newTerm, newSize, newProvenance)\\
                }
            }
       }
    }
    \mycomment{Update ancestor (i.e., new top entry) with sizes from popped entry}
    \If{stack is \textbf{not} empty \textbf{and} stack.columns $>$ 1}
    {
       \For{i=0 to stack.columns}
       {
            \If{popped[i].size+1 $<$ stack.topRow[i].size}
            {
                stack.topRow[i].size $\gets$ popped[i].size+1\\
                stack.topRow[i].provenance $\gets$ \{lastStep(stack.dewey)\}\\
            }
       }
    }
    removeLastDeweyStep(stack.dewey)\\
}   
\label{alg:lcasz}
\caption{CohesiveLCA}
\end{algorithm}

\begin{tabular}{l l l}
\hspace{2cm} $Q$ & $\rightarrow$ & $\ (k)\ |\ T$ \\
\hspace{2cm} $T$ & $\rightarrow$ & $\ (S\ S)$\\
\hspace{2cm} $S$ & $\rightarrow$ & $\ S\ S\ |\ T\ |\ k$\\
\end{tabular}

We now move to define the semantics of cohesive keyword queries. Keyword queries are embedded into data trees. In order to define query answers, we need to introduce the concept of query embedding. In cohesive keyword queries, $m$ occurrences of the same keyword in a query are embedded to one or multiple instances of this keyword as long as these instances collectively contain at least $m$ times this keyword. Cohesive keyword queries may also contain terms, which, as mentioned before, express a cohesiveness relationship on their keyword occurrences. In tree structured data, the keyword instances in the data tree (which are nodes) are represented by their LCA \cite{DBLP:conf/icde/SchmidtKW01, DBLP:conf/sigmod/GuoSBS03, DBLP:journals/www/LiuC11}. The instances of the keywords of a term in the data tree should be indivisible. That is, from the point of view of the instance of a keyword which is external to a term, the subtree rooted at the LCA of the instances of the keywords which are internal to the term is a {\em black box}. Therefore, for a given query embedding, the LCA of the instance of a keyword $k$ which is not in a term $T$ and the instance of a keyword in $T$ should be the same as the LCA of the instance of $k$ and the LCA $l$ of the instances of {\em all} the keywords in $T$ and different than $l$. 

As an example, consider query $Q_1$ =\texttt{(XML keyword search (Paul Cooper) (Mary Davis))} issued against the data tree $D_1$ of Figure~\ref{fig:exampletree2}. In Figure~\ref{fig:exampletree2}, keyword instances are shown in bold and the instances of the keywords of a term below the same article are encircled and shaded. The mapping that assigns \texttt{Paul} to node 8, \texttt{Mary} and \texttt{Cooper} to node 9 and \texttt{Davis} to node 10 is not an embedding since from the point of view of \texttt{Mary}, the subtree rooted at article node 6, which is the LCA of authors 8 and 9 (the instances of \texttt{Paul} and \texttt{Cooper}, respectively), is not a black box: the instance of \texttt{Mary} (article node 9) is part of this subtree. These ideas are formalized below.


\begin{figure*}[t]
\centering
{
\includegraphics[height=0.27\textwidth, keepaspectratio=true]{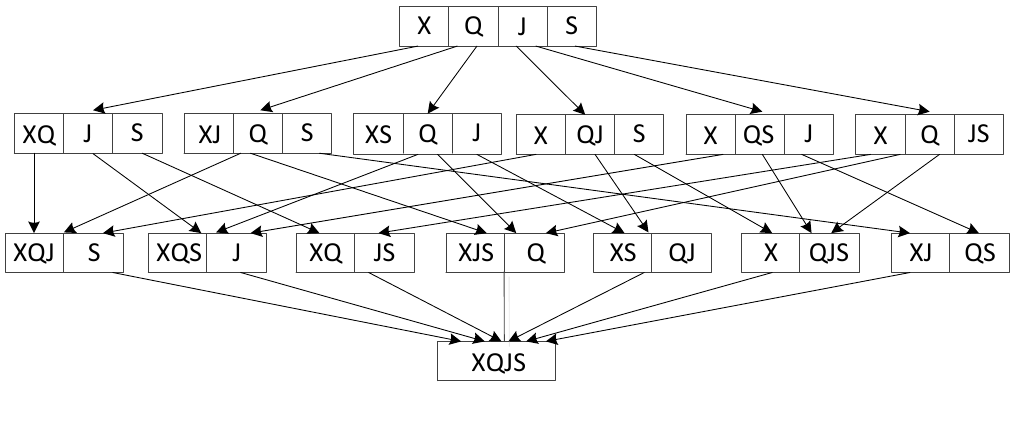}
\label{fig:lattice0}
} 
\caption{Lattice of keyword partitions for the query \texttt{(XML Query John Smith)}}
\label{fig:lattice}
\end{figure*}


\begin{figure*}[!ht]
\centering
{
\subfloat[\scriptsize{(XML\,Query\,(John\,Smith))}]
{
\includegraphics[width=0.24\textwidth, keepaspectratio=true]{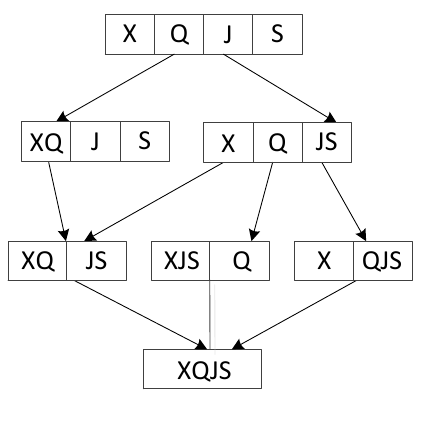}
\label{fig:lattice1}
}
\hspace{1cm}
\subfloat[\scriptsize{((XML\,Query)\,(John\,Smith))}]
{
\includegraphics[width=0.33\textwidth, keepaspectratio=true]{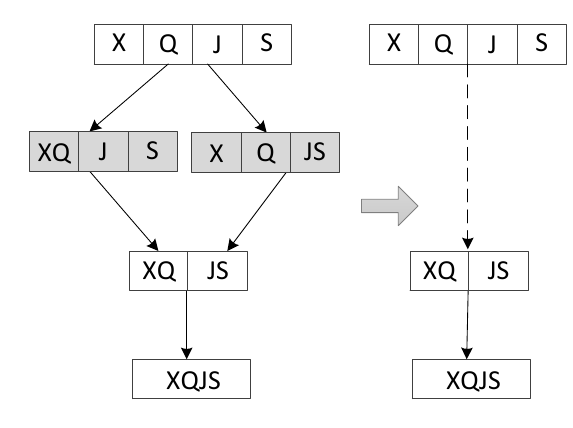}
\label{fig:lattice2}
}
}
\caption{Lattices of keyword partitions for the query keywords \texttt{XML}, \texttt{Query}, \texttt{John} and \texttt{Smith} with different cohesiveness relationships}
\label{fig:componentlattices}
\end{figure*}

\begin{mydef}[Query embedding]
Let $Q$ be a keyword query on a data tree $D.$  An \emph{embedding} of $Q$ to $D$ is a function $e$ from every keyword occurrence in $Q$ to an instance of this keyword in $D$ such that:
\begin{list}{}{\setlength{\leftmargin}{.4cm}\setlength{\parsep}{0cm}
\setlength{\partopsep}{0cm}\setlength{\itemsep}{0.1cm}\setlength{\parskip}{0cm}
\setlength{\labelwidth}{.4cm}
\setlength{\topsep}{0.1cm}}
\item[a.] if $k_1, \ldots, k_m$ are distinct occurrences of the same keyword $k$ in $Q$ and $e(k_1)=\ldots=e(k_m) = n$, then node $n$ contains keyword $k$ at least $m$ times.  
\item[b.] if $k_1, \ldots, k_n$ are the keyword occurrences of a term $T$, $k$ is a keyword occurrence not in $T,$ and $l = lca(e(k_1),\ldots,e(k_n))$ then: (i) $e(k_1)=\ldots=e(k_n)$, or (ii) $lca(e(k), l ) \neq l,\ \forall i \in [1,n]$.
\end{list}
\label{def:embedding2} 
\end{mydef}

Given an embedding $e$ of a query $Q$ involving the keyword occurrences $k_1, \ldots, k_m$ on a data tree $D$, the \emph{minimum connecting tree} (MCT) $M$ of $e$ on $D$ is the minimum subtree of $D$ that contains the nodes $e(k_1),\ldots,e(k_m)$. Tree $M$ is also called an MCT of query $Q$ on $D$. The root of $M$ is the \emph{lowest common ancestor} (LCA) of $e(k_1),\ldots,e(k_m)$ and defines one \emph{result} of $Q$ on $D$. For instance, one can see that the article nodes 2 and 11 are results of the example query $Q_1$ on the example tree $D_1$. In contrast, the article node 6 is not a result of $Q_1$.

We use the concept of LCA size to rank the results in a query answer. Similarly to metrics for flat documents in information retrieval, LCA size reflects the proximity of keyword instances in the data tree. The \emph{size} of an MCT is the number of its edges. Multiple MCTs of $Q$ on $D$ with different sizes might be rooted at the same LCA node $l$. The size of $l$ (denoted $size(l)$) is the minimum size of the MCTs rooted at $l$.

For instance, the size of the result article node 2 of query $Q_1$ on the data tree of $T_1$ is 3 while that of the result article node 11 is 6 (note that there are multiple MCTs of different sizes rooted at node 11 in $D_1$).

\begin{mydef}
The answer to a cohesive keyword query $Q$ on a data tree $D$ is a list $[l_1,\ldots,l_n]$ of the LCAs of $Q$  on $D$ such that $size(l_i) \leq size(l_j), i<j$. 
\end{mydef}

For instance, article node 2 is ranked above article node 11 in the answer of $Q_1$ on $D_1$.

\section{The Algorithm}

We designed algorithm CohesiveLCA for keyword queries with cohesiveness relationships. Algorithm CohesiveLCA computes the results of a cohesive keyword query which satisfy the cohesiveness relationships in the query and are ranked in descending order of their LCA size. The intuition behind CohesiveLCA is that LCAs of keyword instances in a data tree result from combining LCAs of subsets of the same instances (i.e., partial LCAs of the query) bottom-up way in the data tree. CohesiveLCA progressively combines partial LCAs to eventually return full LCAs of instances of all query keywords higher in the data tree. During this process, LCAs are grouped based on the keywords contained in their subtrees. The members of these groups are compared among each other in terms of their size. CohesiveLCA exploits a lattice of partitions of the query keywords.

\vspace*{1.0ex}\noindent{\bf The lattice of keyword partitions. }
During the execution of CohesiveLCA, multiple stacks are used. Every stack corresponds to a partition of the keyword set of the query. Each stack entry contains one element (partial LCA) for every keyword subset belonging to the corresponding partition. As usual with stack based algorithms for processing tree structured data, stack entries are pushed and popped during the computation according to a preorder traversal of the data tree. Dewey codes are exploited to index stack entries which at any point during the execution of the algorithm correspond to a node in the data tree. Consecutive stack entries correspond to nodes related with parent-child relationships in the data tree.

The stacks used by algorithm CohesiveLCA are naturally organized into a lattice, since the partitions of the keyword set (which correspond to stacks) form a lattice. Coarser partitions can be produced from finer ones by combining two of their members. Partitions with the same number of members belong to the same coarseness level of the lattice. Figure~\ref{fig:lattice} shows the lattice for the keyword set of the query \texttt{(XML\,Query\,John\,Smith)}. CohesiveLCA combines partial LCAs following the source to sink paths in the lattice as shown in Figure~\ref{fig:lattice}. 



\vspace*{1.0ex}\noindent{\bf Reducing the dimensionality of the lattice. }
The lattice of keyword partitions for a given query consists of all possible partitions of query keywords. The partitions reflect all possible ways that query keywords can be combined to form partial and full LCAs. Cohesiveness relationships restrict the ways keyword instances can be combined in a query embedding ( Definition~\ref{def:embedding2} ) to form a query result. Keyword instances may be combined individually with other keyword instances to form partial or full LCAs only if they belong to the same term: if a keyword $a$ is ``hidden'' from a keyword $b$ inside a term $T_a$, then an instance of $b$ can only be combined with an LCA of all the keyword instances of $T_a$ and not individually with an instance of $a$. These restrictions result in significantly reducing the size of the lattice of the keyword partitions as exemplified next.

Figure~\ref{fig:componentlattices} shows the lattices of the keyword partitions of two queries. The queries comprise the same set of keywords \texttt{XML}, \texttt{Query}, \texttt{John} and \texttt{Smith} but involve different cohesive relationships. The lattice of Figure~\ref{fig:lattice} is the full lattice of 15 keyword partitions and allows every possible combination of instances of the keywords \texttt{XML}, \texttt{Query}, \texttt{John} and \texttt{Smith}. The query of Figure~\ref{fig:lattice1} imposes a cohesiveness relationship on \texttt{John} and \texttt{Smith}. This modification renders several partitions of the full lattice of Figure~\ref{fig:lattice} meaningless. For instance in Figure~\ref{fig:lattice1}, the partition \texttt{[XJ, Q, S]} is eliminated, since an instance of \texttt{XML} cannot be combined with an instance of \texttt{John} unless the instance of \texttt{John} is already combined with an instance of \texttt{Smith}, as is the case in the partition \texttt{[XJS, Q]}. The cohesiveness relationship on \texttt{John} and \texttt{Smith} reduces the size of the initial lattice from 15 to 7. A second cohesiveness relationship between \texttt{XML} and \texttt{Query} further reduces the lattice to the size of 3, as shown in Figure~\ref{fig:lattice2}. Note that in this case, besides partitions that are not permitted because of the additional cohesiveness relationship (e.g., \texttt{[XJS, Q]}), some partitions may not be productive, which makes them useless. \texttt{[XQ, J, S]} is one such partition. The only valid combination of keyword instances that can be produced from this partition is \texttt{[XQ, JS]}, which is a partition that can be produced directly from the source partition \texttt{[X, Q, J, S]} of the lattice. The same holds also for the partition \texttt{[X, Q, JS]}. Thus, these two partitions can be eliminated from the lattice.


\begin{figure*}[t]
\centering
\subfloat[\scriptsize{Component lattices for terms: (i)\texttt{(XML Keyword Search)}, (ii)\texttt{(Paul Cooper)}, (iii)\texttt{(Mary Davis)} and (iv)\texttt{((XML Keyword Search)(Paul Cooper)(Mary Davis))}}]
{
\includegraphics[width=0.7\textwidth, keepaspectratio=true]{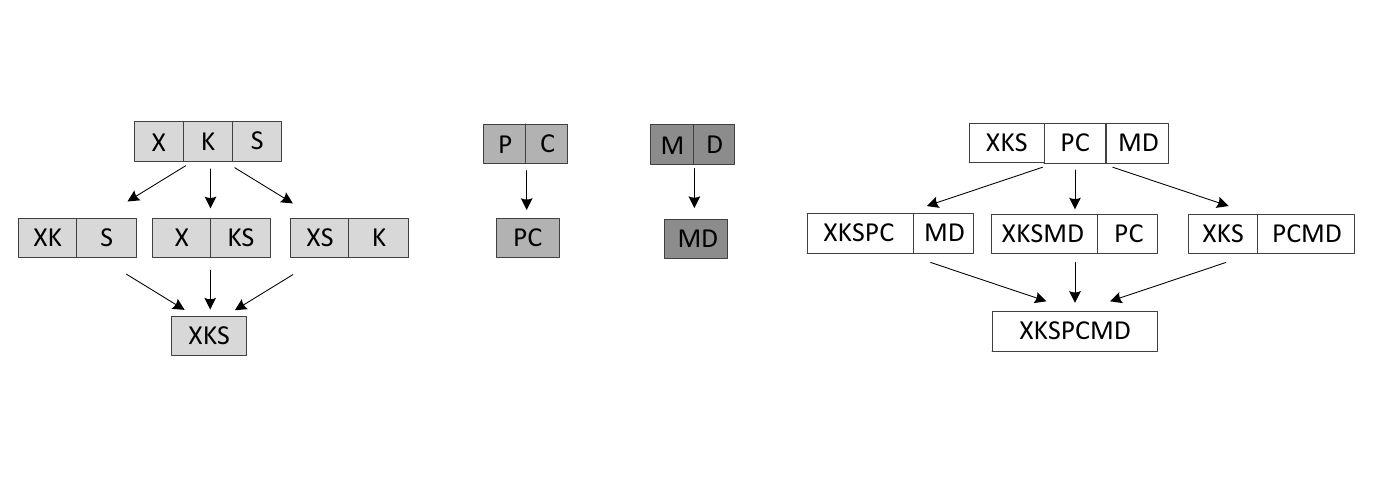}
\label{fig:component_lattices}
} 
\subfloat[\scriptsize{Final lattice}]
{
\includegraphics[width=0.3\textwidth, keepaspectratio=true]{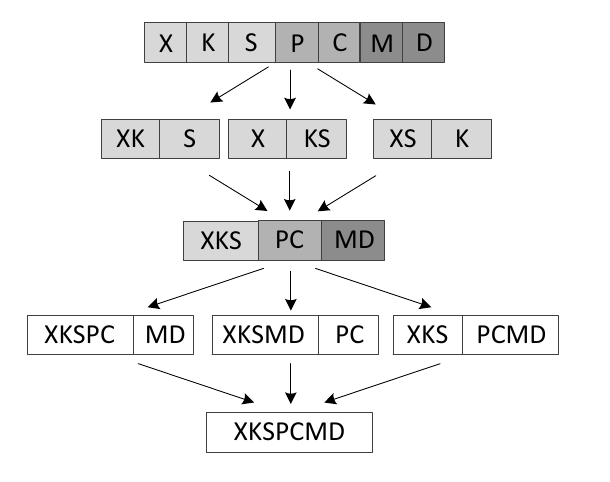}
\label{fig:full_lattice}
}
\caption{Component and final lattices for the query \texttt{((XML Keyword Search) (Paul Cooper) (Mary Davis)))}}
\label{fig:latticeconstruction}
\end{figure*}

\vspace*{1.0ex}\noindent{\bf Algorithm description. }
Algorithm CohesiveLCA accepts as input a cohesive keyword query and the inverted lists of the query keywords and returns all LCAs which satisfy the cohesiveness relationships of the query, ranked on their LCA size.

\begin{function}[t]
\SetKwBlock{simpleblock}{}{}
\SetKwComment{mycomment}{/* }{ */}
\SetAlFnt{\tiny\sf}
\SetNoFillComment
\SetAlgoSkip{}
\small
\textbf{buildLattice}($Q$: query)
\simpleblock
{
   singletonTerms $\gets$ \{keywords(Q)\}\\   
   stacks.add(createSourceStack(singletonTerms)) \mycomment{create the source stack from all singleton keywords}
    constructControlSet($Q$)\mycomment{produce control sets of combinable terms}
    \For{every control set $cset$ in $controlSets$ with not only singleton keywords}
    {stacks.add(createSourceStack(cset))}
     \For{every $s$ in $stacks$}
    {buildComponentLattice($s$)}
    
}

\textbf{constructControlSet}($qp$: query subpattern)
\simpleblock
{
 c $\gets$ new Set()\\
 \For{every singleton keyword $k$ in $s$}
 { c.add($k$)}  
 \For{every subpattern $sqp$ in  $s$}
 { 
	subpatternTerm $\gets$ constructControlSet(sqp)\\	
 	c.add(subpatternTerm)
 }
 controlSets.add(c)\\  
 return newTerm(c) \mycomment{return the cohesive term consisting of component keywords and terms}
}

\textbf{buildComponentLattice}($s$: stack)
\simpleblock
{
 \For{every pair $t1$, $t2$ of terms in $s$}
 {
 	newS $\gets$ newStack(s, t1, t2)\mycomment{construct new stack from s by combining t1 and t2 columns of s}
 	buildComponentLattice(newS)
 }
}

\label{alg:buildLattice}
\vspace*{-1ex}
\caption{buildLattice()}
\end{function}

The algorithm begins by building the lattice of stacks needed for the cohesive keyword query processing (line 2). This process will be explained in detail in the next paragraph. After the lattice is constructed, an iteration over the inverted lists (line 3) pushes all keyword instances into the lattice in Dewey code order starting from the source stack of the lattice, which is the only stack of coarseness level 0. For every new instance, a round of sequential processing of all coarseness levels is initiated (lines 6-9). At each step, entries are pushed and popped from the stacks of the current coarseness level. Each stack has multiple columns corresponding to and named by the keyword subsets of the relevant keyword partition. Each stack entry comprises a number of elements one for every column of the stack. Popped entries contain partial LCAs that are propagated to the next coarseness levels. An entry popped from the sink stack (located in the last coarseness level) contains a full LCA and constitutes a query result. After finishing the processing of all inverted lists, an additional pass over all coarseness levels empties the stacks producing the last results (line 10).

Procedure push() pushes an entry into a stack after ensuring that the top stack entry corresponds to the parent of the partial LCA to be pushed (lines 11-16). This process triggers pop actions of all entries that do not correspond to ancestors of the entry to be pushed into. Procedure pop() is where partial and full LCAs are produced (lines 17-34). When an entry is popped, new LCAs are formed (lines 21-28) and the parent entry of the popped entry is updated to incorporate partial LCAs that come from the popped child entry (lines 29-34). The construction of new partial LCAs is performed by combining LCAs stored in the same entry. 

\vspace*{1.0ex}\noindent{\bf Construction of the lattice. }
The key feature of CohesiveLCA is the dimensionality reduction of the lattice which is induced by the cohesiveness relationships of the input query. This reduction, as we also show in our experimental evaluation, has a significant impact on the efficiency of the algorithm. Algorithm CohesiveLCA does not na\"ively prune the full lattice to produce a smaller one, but wisely constructs the lattice needed for the computation from smaller component sublattices. This is exemplified in Figure~\ref{fig:latticeconstruction}.

Consider the data tree depicted in Figure~\ref{fig:exampletree2} and the query \texttt{((XML Keyword Search) (Paul Cooper) (Mary Davis)))} \linebreak
issued on this data tree. If each term is treated as a unit, a lattice of the partitions of three  items is needed for the evaluation of the query. This is lattice (iv) of Figure~\ref{fig:component_lattices}. Howerver, the input of this lattice consists of combinations of keywords and not of single keywords. These three combinations of keywords each defines its own lattice shown in the left side of Figure~\ref{fig:component_lattices} (lattices (i), (ii) and (iii)). The lattice to be finally used by the algorithm CohesiveLCA is produced by composing lattices (i), (ii) and (iii) with lattice (iv) and is shown in Figure~\ref{fig:full_lattice}. This is a lattice of only 9 nodes, whereas the full lattice for 7 keywords has 877 nodes.

Function~buildLattice() constructs the lattice for evaluating a cohesive keyword query. This function calls another function~buildComponentLattice() (line 8). Function~buildComponentLattice() (lines 18-21) is a recursive and builds all lattices for all terms which may be arbitrarily nested. The whole process is controlled by the $controlSets$ variable which stores the keyword subsets admissible by the input cohesiveness relationships. This variable is constructed by the procedure constructControlSet() (lines 9-17).

\subsection{Algorithm analysis}
\label{sec:complexity}

Algorithm CohesiveLCA processes the inverted lists of the keywords of a query exploiting the cohesiveness relationships to limit the size of the lattice of stacks used.  The size of a lattice of a set of keywords with $k$ keywords is given by the Bell number of $k$, $B_k$, which is defined by the recursive formula: 
$$
B_{n+1} = \sum_{i=0}^{n}{n \choose i} B_i,\ B_0 = B_1 = 1
$$
In a cohesive query containing $t$ terms the number of sublattices is $t+1$ counting also the sublattice of the query (outer term). The size of the sublattice of a term with cardinality $c_i$ is $B_{c_i}$. A keyword instance will trigger in the worst case an update to all the stacks of all the sublattices of the terms in which the keyword participates. If the maximum nesting depth of terms in the query is $n$ and the maximum cardinality of a term or of the query itself is $c$, then an instance will trigger O($nB_c$) stack updates. For a data tree with depth $d$, every processing of a partial LCA by a stack entails in the worst case $d$ pops and $d$ pushes, i.e., O($d$). Every pop from a stack with $c$ columns implies in the worst case $c(c-1)/2$ combinations to produce partial LCAs and $c$ size updates to the parent node, i.e., O($c^2$). Thus, the time complexity of CohesiveLCA is given by the formula:
$$
 O(dnc^2B_c\sum_{i=1}^c{|S_i|})
$$
where $S_i$ is the inverted list of the keyword $i$.
The maximum term cardinality for a query with a given number of keywords depends on the number of query terms. 
It is achieved by the query when all the terms contain one keyword and one term with the exception of the innermost nested term which contains two keywords. Therefore, the maximum term cardinality is $k-t-1$ and the maximum nesting depth is $t$.
Thus, the complexity of CohesiveLCA is:
$$
 O(dt(k-t-1)^2B_{k-t-1}\sum_{i=1}^{k}{|S_i|})
$$
This is a paremeterized complexity which is linear to the size of the input (i.e., $\sum{|S_i|}$) for a constant number of keywords and terms.

\section{Experimental evaluation}

We experimentally studied the effectiveness of the CohesiveLCA semantics and the efficiency of the CohesiveLCA algorithm. 

The experiments were conducted on a computer with a 1.8GHz dual core Intel Core i5 processor running Mac OS Lion. The code was implemented in Java.

We used the real datasets DBLP\footnote{\small{http://www.informatik.uni-trier.de/~ley/db/}}, NASA\footnote{\small{http://www.cs.washington.edu/research
\\
\hspace{1.0cm}/xmldatasets/www/repository.html}} and PSD \footnote{small{http://pir.georgetown.edu/}}and the benchmark auction dataset XMark\footnote{\small{http://www.xml-benchmark.org}}. These datasets display various characteristics. Table~\ref{tab:datasets} shows their statistics. 
\begin{figure}[!h]
\centering
\resizebox{0.48\textwidth}{!}
{
\small
\begin{tabular}{|l|r|r|r|r|}
\hline
 & \emph{DBLP} & \emph{XMark} & \emph{NASA} & \emph{PSD}\\
\hline \hline
size & 1.15 GB & 116.5 MB & 25.1 MB & 683 MB \\
maximum depth & 5 & 11 & 7 & 6 \\
\# nodes & 34,141,216 & 2,048,193 & 530,528 & 22,596,465 \\
\# keywords & 3,403,570 & 140,425 & 69,481 & 2,886,921 \\
\# distinct labels & 44 & 77 & 68 & 70 \\
\# distinct label paths & 196 & 548 & 110 & 97 \\
\hline
\end{tabular}
\captionof{table}{\small{DBLP, XMark, NASA and PSD datasets' statistics}}
\label{tab:datasets}
}
\end{figure}
The DBLP is the largest and XMark the deepest dataset. For the efficiency evaluation, we used the DBLP, XMark and NASA datasets in order to test our algorithm on data with different structural and size characteristics. For the effectiveness experiments, we use dthe largest real datasets DBLP and PSD. The keyword inverted lists of the parsed datasets were stored in a MySQL database. 

\subsection{Efficiency of the CohesiveLCA algorithm}


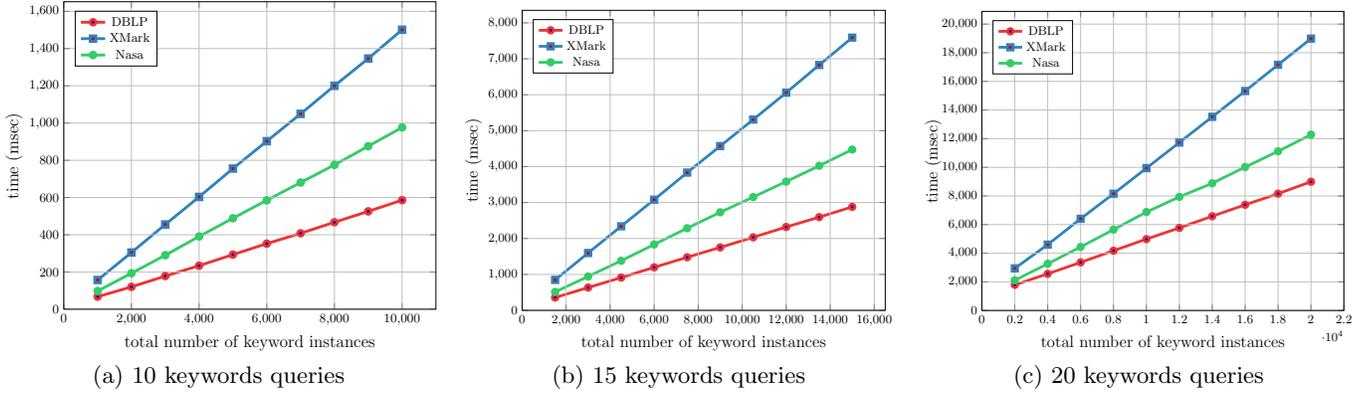
\begin{figure*}[!ht]
\subfloat[10 keywords queries]
{
\resizebox{0.333\textwidth}{!}
{
\begin{tikzpicture}
\begin{axis}[ scale only axis, scaled x ticks = false, scaled y ticks = false, axis on top, height=0.43\textwidth, width=0.52\textwidth, grid=major, xlabel=\large{total number of keyword instances}, ylabel=\large{time (msec)}, ymin=0, xmin=0, legend pos=north west]
\addplot +[myred1, line width=2pt] table [x=instances,y=dblp]{10kws_scaling.dat};
\addplot +[myblue2, line width=2pt] table [x=instances,y=xmark]{10kws_scaling.dat};
\addplot +[mygreen2, line width=2pt] table [x=instances,y=nasa]{10kws_scaling.dat};
\legend{DBLP, XMark, Nasa}
\end{axis}
\end{tikzpicture}
}
\label{fig:CLCAszscalingLargeDBLP}
}
\subfloat[15 keywords queries]
{
\resizebox{0.333\textwidth}{!}
{
\begin{tikzpicture}
\begin{axis}[ scale only axis, scaled x ticks = false, scaled y ticks = false, axis on top, height=0.43\textwidth, width=0.52\textwidth, grid=major, xlabel=\large{total number of keyword instances}, ylabel=\large{time (msec)}, ymin=0, xmin=0, legend pos=north west]
\addplot +[myred1, line width=2pt] table [x=instances,y=dblp]{15kws_scaling.dat};
\addplot +[myblue2, line width=2pt] table [x=instances,y=xmark]{15kws_scaling.dat};
\addplot +[mygreen2, line width=2pt] table [x=instances,y=nasa]{15kws_scaling.dat};
\legend{DBLP, XMark, Nasa}
\end{axis}
\end{tikzpicture}
}
\label{fig:CLCAszscalingXmark}
}
\subfloat[20 keywords queries]
{
\resizebox{0.333\textwidth}{!}
{
\begin{tikzpicture}
\begin{axis}[ scale only axis, scaled x ticks = true, scaled y ticks = false, axis on top, height=0.43\textwidth, width=0.52\textwidth, grid=major, xlabel=\large{total number of keyword instances}, ylabel=\large{time (msec)}, ymin=0, xmin=0, legend pos=north west]
\addplot +[myred1, line width=2pt] table [x=instances,y=dblp]{20kws_scaling.dat};
\addplot +[myblue2, line width=2pt] table [x=instances,y=xmark]{20kws_scaling.dat};
\addplot +[mygreen2, line width=2pt] table [x=instances,y=nasa]{20kws_scaling.dat};
\legend{DBLP, XMark, Nasa}
\end{axis}
\end{tikzpicture}
}
\label{fig:CLCAszscalingNasa}
}
\caption{Performance of CohesiveLCA for queries with 20 keywords varying the number of instances}
\label{fig:CLCAszScaling}
\end{figure*}

In order to study the efficiency of our algorithm we used collections of queries with 10, 15 and 20 keywords issued against the DBLP, XMark and NASA datasets. For each query size, we formed 10 cohesive query patterns. Each pattern involves a different number of terms of different cardinalities nested in various depths. For instance, a query pattern for a 10-keyword query is \texttt{(xx((xxxx)(xxxx))}). We used these patterns to generate keyword queries on the three datasets. The keywords were chosen randomly. In order to stress our algorithm, they were selected among the most frequent ones. In particular, for each  pattern we generated 10 different keyword queries and we calculated the average of their evaluation time. In total, we generated 100 queries for each dataset. For each query, we run experiments scaling the size of each keyword inverted list from 100 to 1000 instances with a step of 100 instances.

\vspace*{1.0ex}\noindent{\bf Performance scalability on dataset size.}
Figure~\ref{fig:CLCAszScaling} shows how the computation time of CohesiveLCA scales when the total size of the query keyword inverted lists grows. Each plot corresponds to a different query size (10, 15 or 20 keywords) and displays the performance of CohesiveLCA on the three datasets. Each curve corresponds to a different dataset and each point in a curve represents the average computation time of the 100 queries that conform to the 10 different patterns of the corresponding query size.  Since the keywords are randomly selected among the most frequent ones the total size of the inverted lists reflects the size of the dataset.

All plots clearly demonstrate that the computation time of CohesiveLCA is linear on the dataset size. This pattern is followed, in fact, by each one of the 100 contributing queries. In all cases, the evaluation times on the different datasets confirm the dependence of the algorithm's complexity on the maximum depth of the dataset: the evaluation on DBLP (max depth 5) is always faster than on NASA (max depth 7) which in turn is faster than on XMark (max depth 11). 

It is interesting to note that our algorithm achieves interactive computation times even with multiple keyword queries and on large and complex datasets.
For instance, a query with 20 keywords and 20,000 instances needs only 20~sec to be computed on the XMark dataset. These results are achived on a prototype without the optimizations of a commercial keyword search system.  To the best of our knowledge, there is no other experimental study in the relevant literature that considers queries of such sizes.


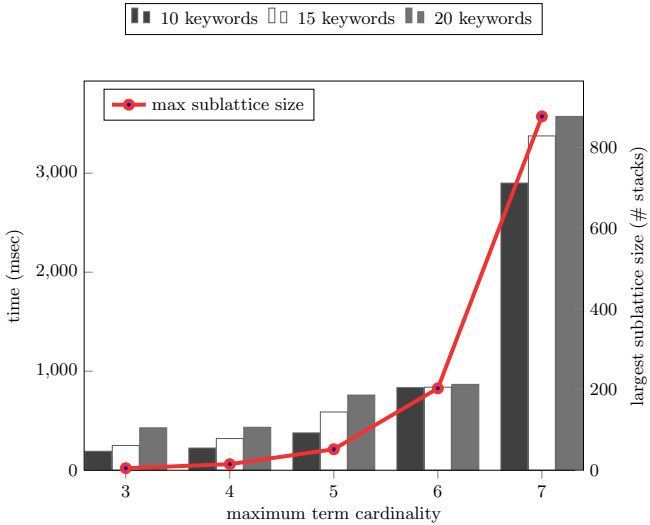
\begin{figure}[!hb]
\resizebox{0.50\textwidth}{!}  %
{
\begin{tikzpicture}
\begin{axis}[ scale only axis,  axis on top, xlabel=maximum term cardinality, ybar, height=0.39\textwidth, width=0.50\textwidth, xtick=data, ylabel=time (msec), tick pos = left, ymin=0, legend style={anchor=north, at={(0.5,1.2)}, column sep=3.2pt}, legend columns=6, bar width=0.027\textwidth, ybar = 0.0005\textwidth]
\addplot +[black!60,fill=black!75] table [x=maxcardinality,y=10kws]{max_cardinality.dat};
\addplot +[black!65,fill=white,postaction={pattern= custom north west lines, pattern color=black!60, hatchspread=1.5pt, hatchthickness=0.5pt}] table [x=maxcardinality,y=15kws]{max_cardinality.dat};
\addplot +[black!60,fill=black!55] table [x=maxcardinality,y=20kws]{max_cardinality.dat};
\legend{10 keywords, 15 keywords, 20 keywords}
\end{axis}
\begin{axis}[ scale only axis, axis on top, height=0.39\textwidth, width=0.50\textwidth, ylabel=largest sublattice size (\# stacks), ylabel near ticks, tick pos = right, ymin=0, axis y line*=right, axis x line=none,  legend style={anchor=north, at={(0.25,0.98)}}]
\addplot +[myred1, line width=2pt] table [x=maxcardinality,y=stacks]{max_cardinality.dat};
\legend{max sublattice size}
\end{axis}
\end{tikzpicture}
}
\label{fig:maxcardinality}
\caption{Performance of CohesiveLCA on queries with 6000 keyword instances for different maximum term cardinalities on the DBLP dataset}
\label{fig:maxcardinality}
\end{figure}

\vspace*{1.0ex}\noindent{\bf Performance scalability on max term cardinality.}
As we showed in \ the analysis of algorithm CohesiveLCA (Section~\ref{sec:complexity}), the key factor which determines the algorithm's performance is the maximum term cardinality in the input query. The maximum term cardinality determines the size of the largest sublattice contributed by the corresponding term(s)  in the construction of the lattice ultimately used by the algorithm. This property is confirmed in Figure~\ref{fig:maxcardinality}. Each bar shows the computation time of queries of a query size with 6000 keyword instances on the DBLP dataset.  The x axis shows the maximum term cardinality of the queries. The computation time shown on the left y axis is averaged over all the queries of a query size with the specific maximum cardinality. The curve shows the evolution of the size of the largest sublattice as the maximum term cardinality increases. The size of the sublattice is indicated by the number of the stacks it contains (right y axis). 

Notice the importance of the maximum cardinality. It is interesting to observe that the computation time depends primarily on the maximum term cardinality and to a much lesser extent on the total number of keywords.  For instance, a query of 20 keywords with a term with maximum cardinality 6 is computed much faster than a query with 10 keywords with maximum cardinality 7. This observation shows that as long as the terms involved are not huge, CohesiveLCA is able to efficiently compute queries with a very large number of keywords.


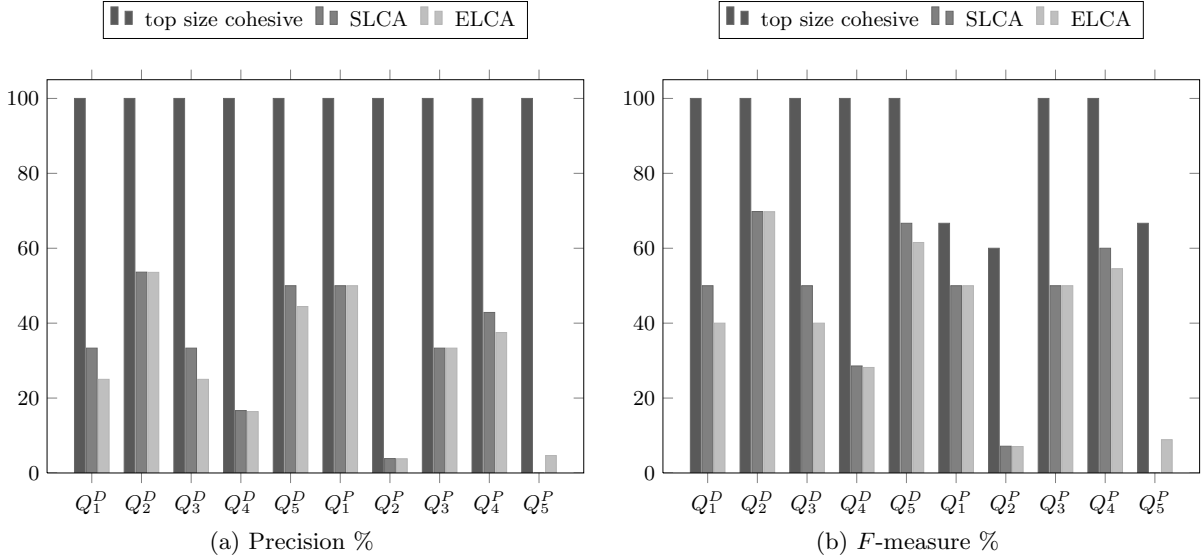
\begin{figure*}[!ht]
\subfloat[Precision \%]
{
\resizebox{0.45\textwidth}{!}  %
{
\begin{tikzpicture}
\begin{axis}[ scale only axis,  axis on top, ybar, height=0.33\textwidth, width=0.45\textwidth, xtick=data, xticklabels={$Q_1^D$,$Q_2^D$,$Q_3^D$,$Q_4^D$,$Q_5^D$,$Q_1^P$,$Q_2^P$,$Q_3^P$,$Q_4^P$,$Q_5^P$}, ymax=105, ymin=0, legend style={anchor=north, at={(0.5,1.2)}, column sep=3.2pt}, legend columns=4, bar width=0.009\textwidth, ybar = 0.001\textwidth]
\addplot +[black!60,fill=black!65] table [y=top1cohesive]{precision.dat};
\addplot +[black!60, fill=black!50, postaction={pattern=custom north west lines, pattern color=white, hatchspread=3pt, hatchthickness=0.5pt}] table [y=SLCA]{precision.dat};
\addplot +[black!30,fill=black!25] table [y=ELCA]{precision.dat};
\legend{top size cohesive, SLCA, ELCA}
\end{axis}
\end{tikzpicture}
}
\label{fig:precision}
}
\subfloat[$F$-measure \%]
{
\resizebox{0.45\textwidth}{!}
{
\begin{tikzpicture}
\begin{axis}[ scale only axis,  axis on top, ybar, height=0.33\textwidth, width=0.45\textwidth, xtick=data, xticklabels={$Q_1^D$,$Q_2^D$,$Q_3^D$,$Q_4^D$,$Q_5^D$,$Q_1^P$,$Q_2^P$,$Q_3^P$,$Q_4^P$,$Q_5^P$}, ymax=105, ymin=0, legend style={anchor=north, at={(0.5,1.2)}, column sep=3.2pt}, legend columns=4, bar width=0.009\textwidth, ybar = 0.001\textwidth]
\addplot +[black!60,fill=black!65] table [x=query,y=top1cohesive]{fmeasure.dat};
\addplot +[black!60, fill=black!50, postaction={pattern=custom north west lines, pattern color=white, hatchspread=3pt, hatchthickness=0.5pt}] table [x=query,y=SLCA]{fmeasure.dat};
\addplot +[black!30,fill=black!25] table [y=ELCA]{fmeasure.dat};\legend{top size cohesive, SLCA, ELCA}
\end{axis}
\end{tikzpicture}
}
\label{fig:fmeasure}
}
\caption{Precision and $\mathcal{F}$-measure of top size Cohesive LCA, SLCA and ELCA filtering semantics}
\label{fig:effectivenessplots}
\end{figure*}

\subsection{Effectiveness of CohesiveLCA semantics}

For our effectiveness experiments we used the real datasets DBLP and PSD. Table~\ref{tab:datasets} lists the queries we evaluated on each dataset. The relevance of the results to the queries was provided by five expert users. In order to cope with a large number of query results we showed to the users the tree patterns of the query results, which are much less than the total number of results, from which they selected the relevant ones. We compared the CohesiveLCA semantics with the SLCA \cite{DBLP:conf/icde/HristidisPB03,DBLP:conf/sigmod/XuP05,DBLP:conf/www/SunCG07,DBLP:conf/icde/ChenP10} and the ELCA  \cite{DBLP:conf/sigmod/GuoSBS03,DBLP:conf/edbt/XuP08,DBLP:conf/edbt/ZhouLL10} filtering semantics. Similarly to our approach, these semantics ignore the node labels and filter out irrelevant results based on the structural characteristics of the results.  

\begin{figure}[!h]
\centering
\resizebox{0.52\textwidth}{!}
{
\begin{tabular}{|l|l|}
\hline
 \multicolumn{2}{|c|}{DBLP}\\
 \hline
 $Q_1^D$ & (proof (scott theorem)) \\
  $Q_2^D$ &  ((ieee transactions communications) (wireless networks))\\
  $Q_3^D$ &  ((Lei Chen) (Yi Guo))\\
  $Q_4^D$ &  ((wei wang) (yi chen))\\
  $Q_5^D$ &  ((vldb journal) (spatial databases))\\
 \hline
  \multicolumn{2}{|c|}{PSD} \\
 \hline
 $Q_1^P$ & ((african snail) mRNA) \\
  $Q_2^P$ & ((alpha 1) (isoform 3)) \\
  $Q_3^P$ & ((penton protein) (human adenovirus 5)) \\
  $Q_4^P$ & ((B cell stimulating factor) (house mouse)) \\
  $Q_5^P$ &  ((spectrin gene) (alpha 1))\\
\hline
\end{tabular}
\captionof{table}{\small{DBLP and NASA queries for efectiveness experiments}}
\label{tab:precisionqueries}
}
\end{figure}

Table~\ref{tab:resultsnumber} shows the number of results for each query and approach on the DBLP and PSD datasets. The CohesiveLCA approach returns all the results that satisfy the cohesiveness relationships in the query.  Since these relationships are imposed by the user, all other results returned by the other approaches are irrelevant. For instance, for query $Q_5^P$, only 3 results satisfy the cohesiveness relationships of the user, and SLCA returns at least 37 and ELCA at least 40 irrelevant results, respectively. CohesiveLCA further ranks the results on LCA size allowing the user to select results of top-k sizes. Usually, the results of the top size are sufficient as we show below.

\begin{figure}[!hb]
\centering
\resizebox{0.40\textwidth}{!}
{
\begin{tabular}{|l||c|c|c|c|}
\hline
 dataset & query & \multicolumn{3}{|c|}{\# of results}\\
\hline
 & & cohesive & SLCA & ELCA\\
\hline
 DBLP & $Q_1^D$ & 2 & 3 & 4\\
 & $Q_2^D$ &  527 & 981 & 982\\
 & $Q_3^D$ &  2 & 3 & 4\\
 & $Q_4^D$ &  11 & 60 & 61\\
 & $Q_5^D$ &  5 & 8 & 9\\
 \hline
 PSD & $Q_1^P$ & 3 & 2 & 3 \\
 & $Q_2^P$ & 14 & 78 & 79 \\
 & $Q_3^P$ & 2 & 4 & 4\\
 & $Q_4^P$ & 4 & 7 & 8 \\
 & $Q_5^P$ & 3 & 40 & 43 \\
\hline
\end{tabular}
\captionof{table}{\small{Number of results of queries on DBLP and PSD datasets}}
\label{tab:resultsnumber}
}
\end{figure}

In order to compare CohesiveLCA with SLCA and ELCA, which provide filtering semantics, we select and return to the user only the top size results. These are the results with the minimum LCA size. The comparison is based on the widely used \emph{precision (P)}, \emph{recall (R)} and $\mathcal{F}$-measure$=\frac{2 P \times R}{P+R}$ metrics  \cite{DBLP:books/aw/Baeza-YatesR2011}. Figure~\ref{fig:effectivenessplots} depicts the results for the three semantics on the two datasets. Since all approaches demonstrate high recall, we only show the precision and  $\mathcal{F}$-measure results in the interest of space.

The diagram of Figure~\ref{fig:precision} shows that CohesiveLCA largely outperforms the other two approaches in all cases. CohesiveLCA shows perfect precision for all queries on both datasets. It also shows perfect $\mathcal{F}$-measure on the DBLP dataset. Its $\mathcal{F}$-measure on the PSD dataset is lower. This is due to the following reason: contrary to the shallow DBLP dataset, the PSD dataset is deep and complex and produces results of various sizes for most of the queries.  Some of the relevant results are not of minimum size and they are missed by top size CohesiveLCA. Nevertheless, all the recall measurements of CohesiveLCA exceed those of the other two approaches. Query $Q_2^P$ highlights the inherent weakness of the SLCA  and ELCA semantics. Not only do they return irrelevant results which do not satisfy cohesiveness relationships (low precision) but they also miss relevant LCAs which satisfy cohesiveness relationships (low recall) because they are ancestors of other relevant LCAs. In the case of $Q_5^P$, the SLCA semantics fails to return any correct result, although it returns in total 40 results (Table~\ref{tab:resultsnumber}), displaying 0\% precision and 0\% recall.

\section{Related work}

Keyword queries facilitate the user with the ease of freely forming queries by using only keywords. Approaches that evaluate keyword queries are currently very popular especially in the web where numerous sources contribute data often with unknown structure and where end users with no specialized skills need to find useful information. However, the imprecision of keyword queries results often in low precision and/or recall of the search systems. Some approaches (a) combine structural constraints \cite{DBLP:conf/vldb/CohenMKS03} with keyword search while others that (b) try to infer useful structural information implied by simple keyword queries \cite{DBLP:conf/icde/BaoLCL09, DBLP:journals/tods/TermehchyW11} by exploiting statistical information of the queries and the underlying datasets. These approaches require a minimum knowledge of the dataset or a heavy dataset preprocessing in order to be able to accurately assess candidate keyword query results.

 The task of locating the nodes in a data tree which most likely match a keyword query has been extensively studied in \cite{DBLP:conf/icde/SchmidtKW01, DBLP:conf/vldb/CohenMKS03, DBLP:conf/icde/HristidisPB03, DBLP:journals/tkde/HristidisKPS06, DBLP:conf/cikm/LiFWZ07, DBLP:conf/sigmod/LiuC07, DBLP:conf/edbt/XuP08, DBLP:conf/edbt/KongGL09, DBLP:conf/edbt/LiLZW10, DBLP:conf/icde/ChenP10, DBLP:conf/sigmod/DimitriouT12, DBLP:journals/tkde/BaoLLC10, DBLP:conf/icde/LiLZW11, DBLP:journals/tkde/LiuWC11, DBLP:journals/tods/TermehchyW11, DBLP:journals/www/NguyenC12, DBLP:conf/dasfaa/AksoyDTW13, DBLP:journals/is/DimitriouTS15}. All these approaches use LCAs of keyword instances as a means to define query answers. The \emph{smallest} LCA (SLCA) semantics  \cite{DBLP:conf/sigmod/XuP05, DBLP:journals/pvldb/LiuC08} validates LCAs that do not contain other descendant LCAs of the same keyword set. A relaxation of this restriction is introduced by \emph{exclusive} LCA (ELCA) semantics \cite{DBLP:conf/sigmod/GuoSBS03, DBLP:conf/edbt/XuP08}, which accepts also LCAs that are ancestors of other LCAs, provided that they refer to a different set of keyword instances.

In a slightly different direction, semantic approaches account also for node labels and node correlations in the data tree. \emph{Valuable} LCAs (VLCAs) \cite{DBLP:conf/vldb/CohenMKS03, DBLP:conf/cikm/LiFWZ07}  and \emph{meaningful} LCAs \cite{DBLP:conf/vldb/LiYJ04} (MLCAs) aim at ``guessing'' the user intent by exploiting the labels that appear in the paths of the subtree rooted at an LCA.  All these semantics are restrictive and depending on the case, they may demonstrate low recall rates as shown in \cite{DBLP:journals/tods/TermehchyW11}.
 
The efficiency of algorithms that compute LCAs as answers to keyword queries depend on the query semantics adopted. By design they exploit the adopted filtering semantics to prune irrelevant LCAs early on in the computation. Stack based algorithms are naturally offered to process tree data. In \cite{DBLP:conf/sigmod/GuoSBS03} a stack-based algorithm that processes inverted lists of query keywords and returns ranked ELCAs was presented. This approach ranks also the query results based on precomputed tree node scores inspired by PageRank \cite{DBLP:journals/cn/BrinP98} and IR style keyword proximity in the subtrees of the ranked ELCAs. In \cite{DBLP:conf/sigmod/XuP05}, two efficient algorithms for computing SLCAs are introduced, exploiting special structural properties of SLCAs. This approach also introduces an extension of the basic algorithm, so that it returns all LCAs by augmenting the set of already computed SLCAs. Another algorithm for efficiently computing SLCAs for both AND and OR keyword query semantics is developed in \cite{DBLP:conf/www/SunCG07}. The Indexed Stack \cite{DBLP:conf/edbt/XuP08} and the Hash Count \cite{DBLP:conf/edbt/ZhouLL10} algorithms improve the efficiency of \cite{DBLP:conf/sigmod/GuoSBS03} in computing ELCAs. Finally, \cite{DBLP:conf/icde/BaoLCL09, DBLP:journals/tkde/BaoLLC10} elaborate on sophisticated ranking of candidate LCAs aiming primarily on effective keyword query answering.

Filtering semantics are often combined with (i) structural and semantic correlations \cite{DBLP:conf/sigmod/GuoSBS03, DBLP:journals/isci/LiLFZ09, DBLP:conf/icde/ChenP10,DBLP:journals/tods/TermehchyW11, DBLP:conf/vldb/CohenMKS03, DBLP:conf/dasfaa/AksoyDTW13}, (ii) statistical measures \cite{DBLP:conf/sigmod/GuoSBS03, DBLP:conf/vldb/CohenMKS03, DBLP:journals/isci/LiLFZ09, DBLP:conf/icde/ChenP10, DBLP:conf/edbt/LiLZW10, DBLP:journals/tods/TermehchyW11} and (iii) probabilistic models \cite{DBLP:journals/tods/TermehchyW11, DBLP:journals/www/NguyenC12, DBLP:conf/icde/LiLZW11} to perform a ranking to the results set. Nevertheless, such approaches require expensive preprocessing of the dataset which makes them impractical in the cases of fast evolving data and streaming applications.

The concept of LCA size for ranking keyword query results was initially introduced in \cite{DBLP:conf/sigmod/DimitriouT12, DBLP:journals/is/DimitriouTS15}. An algorithm that computes all the LCAs ranked on LCA size that exploits a lattice of stacks is also presented in these papers.  However, this algorithm is designed for flat keyword queries and cannot handle cohesive queries, which is the focus of the present work.

\newpage
\section{Conclusion}
Current approaches for assigning semantics to keyword queries on tree data cannot cope efficiently or effectively with the large number of candidate results and produce answers of low quality. This poor performance cannot offset the convenience and simplicity offered to the user by the keyword queries. In this paper we claim that the search systems cannot guess the user intent from the query and the characteristics of the data to produce high quality answers on any type of dataset and we introduce a cohesive keyword query language which allows the users to naturally and effortlessly express cohesiveness relationships on the query keywords. We design an algorithm which builds a lattice of stacks to efficiently compute cohesive keyword queries and leverages cohesiveness relationships to reduce its dimensionality. Our theoretical analysis and experimental evaluation show that it outperforms previous approaches in producing answers of high quality and scales smoothly succeeding to evaluate efficiently queries with a very large number of frequent keywords on large and complex datasets when previous algorithms for flat keyword queries fail. 

We are currently working on alternative ways for defining semantics for cohesive keyword queries on tree data and in particular in defining skyline semantics which considers all the cohesive terms of a query in order to rank the query results.

\balance
\bibliographystyle{abbrv}
\bibliography{kwsearchxml}
\end{document}